\documentclass[12pt,preprint]{aastex}
\usepackage{natbib,times,mathptmx,epsfig}   

\shorttitle{CBI Observations of the Polarization of the CMBR}
\shortauthors{Cartwright et~al.}

\slugcomment{Submitted to Astrophysical Journal on 6 October 2004}

\begin{document}

\title{Limits on the Polarization of the Cosmic Microwave Background Radiation at Multipoles up to $\ell\sim 2000$}

\author{J. K. Cartwright\altaffilmark{1}, T. J. Pearson, A. C. S. Readhead, M. C. Shepherd, and J. L. Sievers\altaffilmark{2}}
\affil{Department of Astronomy, California Institute of Technology}
\affil{1200 East California Blvd, Pasadena, CA 91125}

\author{G. B. Taylor}
\affil{National Radio Astronomy Observatory}
\affil{P.O. Box 0, Socorro, NM 87801}

\altaffiltext{1}{present address:  Kavli Institute for Cosmological Physics, University of Chicago,
	5640 South Ellis Avenue, Chicago, IL 60637}

\altaffiltext{2}{present address:  Canadian Institute for Theoretical Astrophysics, University of Toronto,
	60 St. George Street, Toronto, Ontario, M5S 3H8, Canada}

\begin{abstract}

We report upper limits on the polarization of the CMBR as measured with the 
Cosmic Background Imager, a 13 element interferometer that operates 
in the 26-36 GHz band and is sited on Llano de Chajnantor in northern Chile. 
The array consists of 90-cm Cassegrain antennas mounted on a steerable 
platform that can be rotated about the optical axis to facilitate 
polarization observations.  
The CBI employs single-mode circularly polarized receivers and it samples
multipoles from $\ell\sim 400$ to $\ell\sim 3500$. 
The polarization data were calibrated on 3C279 and Tau A.
The polarization observations consist of 278 hours of data on two fields
taken in 2000, during the first CBI observing season.
A joint likelihood analysis of the two fields yields three upper limits (95\% c.l.) for 
$\mathcal{C}_{\ell}^{EE} = C^{EE}\ell(\ell+1)/2\pi$ under the assumption that $\mathcal{C}_{\ell}^{BB}\equiv 0$: 
49.0 $\mu$K$^2$ $(\ell=603)$; 164 $\mu$K$^2$ $(\ell=1144)$; and 630 $\mu$K$^2$ $(\ell=2048)$.
\end{abstract} 


\section{Introduction}
\label{section:intro}

The Cosmic Microwave Background Radiation (CMBR) provides a unique means of 
testing many aspects of the Standard Model of the early universe.
All variations of the Model agree that the CMBR 
is the redshifted radiation from the initial 
plasma, and that as such, it contains clues about the fundamental 
characteristics of the universe
\citep[e.g.,][]{kamionkowski99,hu02}.
This information resides in the spatial fluctuations of the 
total intensity and polarization of the CMBR.
The past decade has seen the emergence of low noise detector 
technologies that are propelling us into an new era of precision 
measurements of the characteristics of the CMBR.
Observations have established the existence of intensity anisotropies 
with ${\delta T/T_0} \sim10^{-5}$ on scales of 
$\theta\sim$ 0.1-0.5$^{\circ}$, \cite[e.g.,][]{halverson02,netterfield02,kuo02,hinshaw03,readhead04a}.
In contrast to the fluctuations in total intensity, polarization anisotropies are sufficiently 
small to have eluded detection until very recently \cite[]{kovac02,kogut03,leitch04,readhead04b}.

Standard models predict that Thomson scattering at the surface of last 
scattering will polarize the fluctuations at the 10\% level on scales of tens of arcminutes 
\cite[e.g.,][]{kamionkowski97}.
The Cosmic Background Imager (CBI) is one of several experiments that have used the technique of 
radio interferometry to measure the spatial power spectrum of these fluctuations.
Besides CBI, and its sister instrument, the Degree Angular Scale Interferometer (DASI), 
recent experiments have employed a variety of methods to achieve sensitivities that approach 
cosmologically important levels:  
POLAR \cite[]{keating01}, Saskatoon \cite[]{wollack93,netterfield97},
PIQUE \cite[]{hedman01,hedman02}, CAPMAP \cite[]{barkats04}, and ATCA \cite[]{subrahmanyan00}.
The CBI performed preliminary polarization observations in 2000, and the results 
of this work are reported here.
This initial set of observations demonstrated the CBI's polarization capabilities \cite[]{cartwright03}, and,
on the basis of the success of this work, the telescope has been upgraded and dedicated to polarization 
observations since September 2002;
the detections that resulted from these observations are presented in a recent paper by \cite{readhead04b}.
Other ongoing polarization experiments include BOOMERANG \cite[]{masi02}, MAXIPOL \cite[]{johnson03}, 
COMPASS \cite[]{farese03}, QUEST \cite[]{piccirillo02}, WMAP \cite[]{bennett03}, and 
Planck \cite[]{delabrouille02,villa02}, the latter two of which are all-sky satellite 
missions.

\section{The Cosmic Background Imager}

The Cosmic Background Imager is a 13-element interferometer that operates in the 
26-36 GHz band \cite[]{padin02}.
The array consists of 90-cm Cassegrain antennas mounted on a single, fully steerable platform.
The antenna platform employs the standard alt-az axes, as well as a rotational degree of freedom 
about the telescope optical axis;  this latter feature facilitates polarization observations.
The platform allows a range of positions for the telescopes, permitting observations of 
anisotropies on multipoles $\ell\sim 400$--3500;  this range encompasses the scales 
over which standard models predict that much of the power in total intensity and polarization
fluctuations is to be found.
The observations reported in this paper concentrate on the $400<\ell<2400$ region, to 
which the CBI is particularly well-matched.

The CBI employs single-mode circularly polarized receivers.
In these initial pioneering polarization observations with the CBI the 
main focus was to determine the suitability of the instrument for 
polarization studies and to understand the primary sources of systematic error.
To implement a polarization detection effort in parallel with the 
intensity observations that constituted the CBI's primary mission, 
we configured 12 receivers for {\em LCP} and one receiver for {\em RCP};  the 
resulting array consisted of 66 total intensity ({\em LL}) and 12 cross-polarized 
({\em LR}) baselines, all spanning $400<\ell<3500$.
Each receiver has a quarter-wave plate that determines its 
polarization.
The CBI was configured for polarization observations from January to October 2000,
at which point the single {\em LCP} antenna was converted back to {\em RCP}.

A single interferometer baseline measures a {\em visibility}, which is the 
Fourier transform of the intensity distribution on the sky.
An {\em LCP} and {\em RCP} antenna pair form an {\em LR} baseline, which measures 
the cross-polarized visibility $\mathcal{V}^{LR}$ at a point ${\bf u}=(u,v)$ in the aperture plane:
\begin{equation}
\mathcal{V}^{LR}({\bf u})=\int\int A({\bf x}-{\bf x}_0)[Q({\bf x})-iU({\bf x})]e^{2i\theta}e^{-2\pi i{\bf u}\cdot{\bf x}}d^2{\bf x}
\end{equation}
\noindent $A({\bf x}-{\bf x}_0)$ is the primary beam pattern, which is assumed to be the same for both antennas 
that form the baseline, and is centered at ${\bf x}_0$ on the sky;  $\theta=\tan^{-1}(u/v)$;
and $Q({\bf x})$ and $U({\bf x})$  
are Stokes parameters that describe the distribution of polarized flux on the sky.
Although the integrals are evaluated over the entire sky, $A({\bf x}-{\bf x}_0)$ confines the signal of interest 
to the region of the sky in view of the primary beam.
The interplay between the kernal of the Fourier Transform and the primary beam determines the 
range of angular scales to which the baseline is sensitive;  for an observation at wavelength 
$\lambda$ on a baseline of length $b$, the synthesized beam 
$\theta_s\sim |{\bf u}|^{-1}\sim \lambda/b$ determines the resolution, while the primary beam,
for which $\theta_p^{\rm FWHM} = 46.5^{\prime}(\lambda/{\rm 1\ cm})$, sets the field of view.
\cite{leitch02} discusses the application of this method to polarization observations with DASI.

In the configuration for these observations, the CBI directly 
measures $\mathcal{V}^{LL}({\bf u})$ and $\mathcal{V}^{LR}({\bf u})$.
$I({\bf x})$ can be obtained from measurements of $\mathcal{V}^{LL}({\bf u})$ alone in
the absence of circular polarization, but both $\mathcal{V}^{LR}({\bf u})$
and $\mathcal{V}^{RL}({\bf u})$ are required to obtain $Q({\bf x})$ and $U({\bf x})$.
Although an {\em LR} baseline does not directly measure {\em RL}, we can obtain it via rotation of the 
deck about the optical axis, since $\mathcal{V}^{LR}({\bf u}) = [\mathcal{V}^{RL}(-{\bf u)}]^*$ 
\cite[]{conway69}:  a 
180$^{\circ}$ deck rotation permits determination of $Q({\bf x})$ and $U({\bf x})$.
Although {\em Q} and {\em U} are necessary for imaging, for the likelihood analysis we use $LR$ sampled 
over both halves of the $(u,v)$ plane (Section 4).

The CBI antenna elements are Cassegrain dishes.
Good polarization performance favors a clear aperture, but the Cassegrain optics do not impair the 
performance because the secondary of each telescope is supported by polystyrene feedlegs 
that are transparent at centimeter wavelengths.
Nonetheless, the optics do introduce contamination into the cross-polarized visibilities.
To reduce crosstalk, the antennas are surrounded by cylindrical shields;
unpolarized blackbody emission from the ground is polarized upon scattering from the insides of the 
antenna shields to the antenna feeds.
This spurious spillover signal dominates the cross-polarized visibilities on short baselines, but 
the lead-trail observing technique eliminates this contamination (Section 3).
To test for the presence of spurious off-axis polarization, we measured the instrumental polarization 
at the half-power points of the primary beam in the four cardinal directions.
A $\chi^2$ test demonstrated that the instrumental polarization at the these points is consistent with 
that at the antenna boresight, showing that the instrumental polarization properties do not degrade 
rapidly as one moves off-axis.

\subsection{Polarization Calibration}

The cross-polarized visibilities must be calibrated to measure the gain and instrumental polarization.
To first order, the raw cross-polarized visibility for the baseline using antennas {\em j} and {\em k} 
is given by 

\begin{equation}
\mathcal{V}^{LR}_{jk}({\bf u})= G_{jk}\Big\lbrack{[{\tilde Q}({\bf u})-i{\tilde U}({\bf u})]e^{-2i\psi}+\epsilon_{jk}{\tilde I}({\bf u})}\Big\rbrack
\end{equation}

\noindent where $G_{jk}$ and $\epsilon_{jk}$ denote the {\em baseline}-based instrumental gain and polarization, 
respectively, and ${\tilde I}({\bf u})$, ${\tilde Q}({\bf u})$ and ${\tilde U}({\bf u})$
are the Fourier Transforms of $I({\bf x})$, $Q({\bf x})$ and $U({\bf x})$.
$G_{jk}$ and $\epsilon_{jk}$ are both are complex quantities, and must be evaluated for each of the ten CBI bands.
The instrumental polarization, or {\em leakage}, permits the total intensity to contaminate 
the cross-polarized visibilities;  
for observations of the CMBR, uncorrected leakage will cause measurements of polarization 
fluctuations at a particular $\ell$ to be contaminated by the total intensity fluctions at the same $\ell$. 
$\psi$ denotes the deck orientation about the optical axis.
To determine $G_{jk}$ and $\epsilon_{jk}$, we 
observe a source of known polarization and total intensity at a variety of deck orientations, and the 
change in $\psi$ modulates the first term of Equation 2 relative to the second.
Observations at a minimum of two different deck orientations are required to obtain both $G_{jk}$ 
and $\epsilon_{jk}$.

To calibrate the $LR$ visibilities for the CMBR deep fields, Equation 2 is evaluated at each $(u,v)$ point,
using the values of $G_{jk}$ and $\epsilon_{jk}$ determined from the calibration observations,
together with measurements of ${\tilde I}({\bf u})$, to isolate ${\tilde Q}({\bf u})-i{\tilde U}({\bf u})$.
During the 2000 observing season, the array configuration and observational strategy precluded $LL$ 
matches for all $LR$ visibilities;
for short baselines ($b\sim 100$ cm or $\ell\sim 600$), where we expect the greatest signal, the 
loss of data was not substantial, but for several of the longer baselines ($b\sim 300$ cm or $\ell\sim$ 1900), 
the lack of $LL$ matches precluded the use of all data.  

\begin{figure}
\centerline{\psfig{file=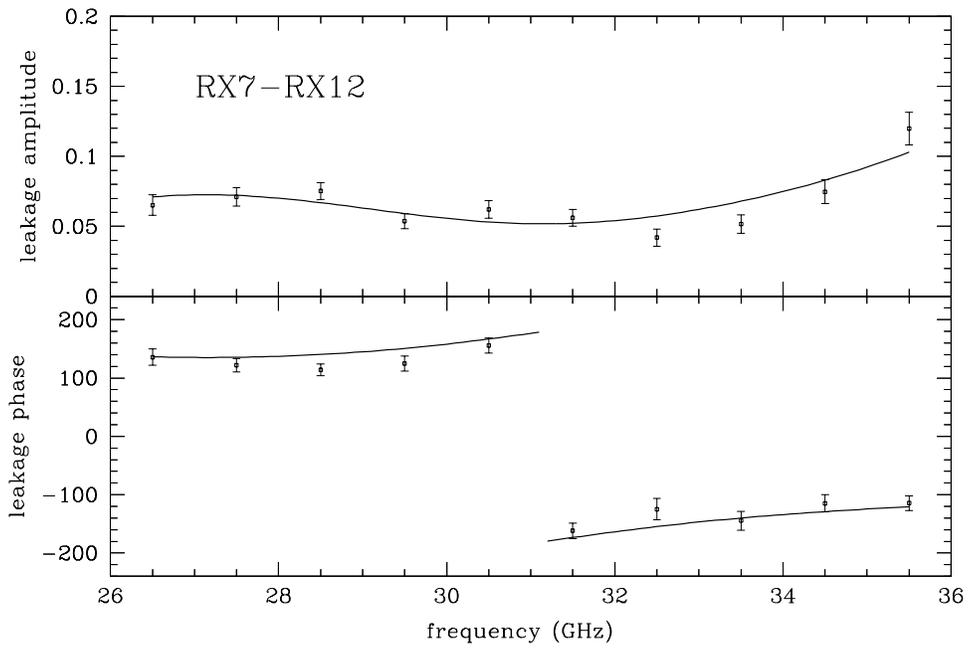,width=5.0in,angle=0}}
\caption[]{Comparison of leakage model fit to leakage data for the baseline between 
RX7 and RX12, across all ten channels.
The upper figure shows the leakage amplitude $\epsilon_A$, in units for 
which 1.0 corresponds to 100\% leakage,
while the lower shows the leakage phase $\epsilon_{\phi}$.
Points represent measurements of the leakage, while lines show the model.}
\label{figure:plot_leakage_model-1}
\end{figure}

In the present observations, the amplitude of the instrumental polarization averages $\sim 8\%$ 
for all baselines and all channels and can approach 20\%.
The receiver was modeled to understand the source of the leakage.
The instrumental polarization is dominated by bandpass errors in the quarter-wave plates;
at the edges of the $\sim30\%$ fractional band, the insertion phase of the quarter-wave 
plates departs from $\lambda/4$ by several percent.
In addition, assembly errors can cause the plate orientation to depart from the ideal 45$^{\circ}$ 
with respect to the rectangular guide that follows it.
A receiver model that incorporates these errors shows excellent agreement with the measured leakage
(Figure~\ref{figure:plot_leakage_model-1}).
The modeled receiver characteristics that give rise to $\epsilon$ are stable, so we infer that 
the measured leakage is stable as well.
High signal-to-noise-ratio measurements of the leakage at regular intervals demonstrated 
that it remained stable over timescales spanning many months.

The polarization data were calibrated with observations of extragalactic radio source 3C279 
and the supernova remnant Taurus A (the Crab nebula).
3C279, a bright extragalactic radio source, served as the primary polarization calibrator.
With $I\sim 25$ Jy and $m=|P|/I\sim 10\%$ at 31 GHz (where $|P|=\sqrt{Q^2+U^2}$), and no 
significant extended emission on CBI scales, 3C279 permits quick calibration observations.
3C279 is variable, however, so it was monitored at monthly intervals throughout the polarization 
campaign with the NRAO Very Large Array\footnote{The National Radio Astronomy Observatory is a 
facility of the National Science Foundation operated under cooperative agreement by 
Associated Universities, Inc.} at 22.46 GHz and 43.34 GHz.
3C279 showed some activity during the January-May period;  at 22.46 GHz, its fractional polarization 
changed by $\sim20\%$ ($\delta m\sim0.02$), while its position angle rotated by 10$^{\circ}$ during the 
same period.
These changes were approximately linear at 22.46 GHz and occasionally discontinuous at 43.34 GHz.
Although the VLA observations yield $I$, $Q$, and $U$, we transferred only the fractional 
polarization {\em m} and the position angle $\chi$ (where 2$\chi=\tan^{-1}(U/Q)$) to the CBI. 
This choice permits us to use daily measurements of 3C279's total 
intensity with the CBI to set the flux density scale for the polarization observations.
The absolute uncertainty of the total intensity calibration is 4\%.

We required two interpolations to apply the VLA values for $m$ and $\chi$ to the CBI observations.
The first interpolation transfers $m$ and $\chi$ from the two VLA channels to the 26-36 GHz band.
Measurements of the total intensity of 3C279 in the ten CBI bands show that the total intensity is 
well characterized by a power law, and in light of this uniform behavior we made a simple linear fit 
to both $m$ and $\chi$.
The statistical uncertainty on this fit is typically less than $5\%$ in amplitude and 3$^{\circ}$ in 
position angle.
The second interpolation transfers {\em m} and $\chi$ from the dates of the VLA observations to the 
intervening CBI observations.
Again, a linear interpolation was used.
While the statistical uncertainty in this interpolation is generally small ($< 5\%$ in amplitude and 
3$^{\circ}$ in position angle) and trivial to compute, 
the systematic uncertainty is harder to estimate, particularly for the 43.34 GHz data;  the changes 
in {\em m} and $\chi$ between VLA 
observations at 43.34 GHz in one interval are quite high ($\sim 20\%$ and $\sim 10^{\circ}$), although 
only $\sim 1\%$ of the CMBR data were taken during this interval.
Regular measurements of the total intensity of 3C279 with the CBI show that it does not 
undergo excursions beyond those seen in the VLA data, however, so we assume that the temporal 
variations in the polarization characteristics do not exceed those in the VLA data.
The measurement uncertainties in the VLA data are typically 3\%, so the uncertainties in 
the interpolated VLA data can be as high as 8\%.

3C279 was observed nearly every night at a pair of deck positions separated by 
90$^{\circ}$;  each observation lasted 5$^m$ and was accompanied by a trailing 
field to measure contamination from ground spillover (Section 3).
The total uncertainty in the 3C279 calibration is typically 9\%, of which 8\% arises from the 
uncertainty in the VLA data, 3\% results from the uncertainty in the CBI $LR$ observations, and
4\% arises from the flux scale, which is set by the uncertainty in the CBI's $LL$ calibration.

Tau A served as the polarization calibrator for nearly 40\% of the polarization data.
Tau A is marginally resolved by the CBI, so we required a simple model for its morphology.
There are no published data on Tau A's polarized emission at 31 GHz, 
so we transferred the calibration on 3C279---obtained directly from a nearly contemporaneous VLA 
observation---to the Tau A observations and derived a model.
Our Tau A model consists of single elliptical Gaussian components for each of $I$, $Q$, and $U$;
these model components are shown in Table~\ref{table:taua_models},
and this simple model is applicable over ranges of $|u|\sim 100$--500 and the 26-36 GHz band.
The spectral indices for the two polarized components were constrained to be 
that of the total intensity: $\alpha = -0.3$,  where $S_{\nu}\propto\nu^{\alpha}$.  

We performed a number of supporting observations to assess the accuracy of the polarization calibration.
We included 3C273 in the VLA monitoring campaign, and observations of 3C273 with the CBI provide a
test of the internal consistency of the polarization calibration.
3C273 is a $\sim 25$ Jy, $\sim 5\%$ polarized source at centimeter wavelengths, and the 
polarization we recover from CBI observations of 3C273 
is consistent with the VLA observations within the statistical and systematic 
uncertainties of the calibration.
Cross-checks on observations of 3C279 provide estimates of the uncertainty on the 
calibration with the Tau A model in Table~\ref{table:taua_models};  using Tau A as 
a calibrator, we recover $m$ and $\chi$ for 3C279 to within $\sim 10\%$ and $\sim 5^{\circ}$, respectively.
We regard these values as the uncertainties on the polarization calibration.

\section{Observations of the CMBR}

The data presented here were obtained from deep observations of two fields, the
08$^h$ field ($\alpha = 08^h44^m40^s$, $\delta = -3^{\circ}10^{\prime}00^{\prime\prime}$), and the
20$^h$ field ($\alpha = 20^h48^m40^s$, $\delta = -3^{\circ}30^{\prime}00^{\prime\prime}$);
measurements of total intensity fluctuations in these fields have been reported by \cite{mason02}.
These fields are a subset of a group of four fields spaced at equal intervals in right 
ascension\footnote{The CBI's elevation limit of 43$^{\circ}$ constrains the 
time on source to $\sim 6^h$ per day.}
that were selected for minimum contamination from diffuse galactic emission.
Both fields are at galactic latitudes greater than 24$^{\circ}$.
Each field is the size of a single beam, or $45^{\prime}$ FWHM at the band center (Section 2).
Simple extrapolations from Haslam 408 MHz maps suggested that for both fields, the polarization 
fluctuations from synchrotron emission at 1 cm on CBI scales would be smaller than the CMBR 
polarization fluctuations \cite[]{haslam82}.

The observations presented in this paper were obtained between January and October 2000.
The 08$^h$ field was observed from January through the end of May, and the 20$^h$ field was 
observed from August through the end of October, at which point the array was  
dedicated to total intensity observations until September 2002.
This work encompassed 99 nights of observations:  44 nights on the 08$^h$ field and 55 nights 
on the 20$^h$ field, which yielded 130$^h$ and 148$^h$ of data, respectively.
The 08$^h$ field observations spanned two array configurations, while the 20$^h$ field was observed with a third.
The weather at the Chajnantor site was generally generally excellent when observations were not precluded 
by snow storms, and less than 1\% of the data were flagged.

The observing strategy was guided by several considerations.
The visibilities measured on the short baselines are contaminated by ground spillover, so we 
observed fields in pairs separated by 8$^m$ in right ascension and 
differenced the pairs offline to reject the common spillover contribution\footnote{The positions given above are 
those of the leading fields.} \cite[]{mason02}.
To within the uncertainties of the visibilities, the $LL$ and $LR$ visibilities show 
no evidence of spillover after differencing.
The observations were performed at night and when the moon was greater than 60$^{\circ}$
from the fields.
Each lead/trail pair was tracked in constant parallactic angle, and, after each pair,
the deck position was advanced by either 20$^{\circ}$ or 30$^{\circ}$.
Each 8$^m$ scan consists of $\sim 50$ 8.4$^s$ integrations;  $\sim 15\%$ of each scan is lost to 
calibrations and slews.

We performed a number of consistency tests on the CMBR data prior to the likelihood analysis.
We first applied a jackknife test to assess the accuracy of the noise estimates.
The visibility uncertainty for each $8^m$ scan was estimated from the scatter in the $\sim 50$ 
integrations in the scan.
For the jackknife test, the real and imaginary visibilities were sorted into two interleaved 
sets corresponding to alternating dates, and at each $(u,v)$ point, each set was averaged over time.
The two sets were the differenced on a point-by-point basis in the $(u,v)$ plane.
We are most concerned about effects on the shortest baselines, for which we expect the greatest signal,
and conversely, for which the spillover contamination is greatest.
To that end, for the 08$^h$ field we computed $\chi^2_{\nu}=1.03$ with $\nu=590$ (probability-to-exceed = 30\%) 
for the real components, and $\chi^2_{\nu}=1.11$ (p.t.e. = 3\%) for the imaginaries.
Similarly, for the 20$^h$ field visibilities ($\nu=720$), 
we find $\chi^2_{\nu}=1.00$ (real) and  $\chi^2_{\nu}=0.97$ (imaginary), both of which are consistent with unity. 

We were also concerned that systematic errors in the polarization of the calibrator sources---particularly 
the Tau A model---would give rise to errors in the calibration of the CMBR data.
Since $\sigma_\mathcal{V}\propto |G_{jk}|$, we used the visibility uncertainties as a proxy 
for the amplitude component of the $LR$ gain calibration;  we averaged the $LR$ visibility uncertainties 
for the CMBR data and compared them to those for $LL$, the gain calibration of which we believe to be 
accurate to $\sim 4\%$.
After accounting for a slight (4\%) excess in system noise for RX12---the orthogonally polarized antenna
that is common to all $LR$ visibilities---we find that $\langle\sigma_{LR}\rangle\sim\langle\sigma_{LL}\rangle$
to within 10\%, which is consistent with the results of the calibration cross-checks discussed in Section 2.1.

\section{Additional Observations of Polarized Sources}

We observed several polarized sources to assess the polarization performance of the CBI.
Centaurus A (NGC 5128) is a nearby active galaxy that exhibits a rich 
variety of polarized structure over a range of angular scales at centimeter wavelengths. 
W44 has several janskies of polarized emission at 1 cm, and its size of $\sim 30^{\prime}$ 
is comparable to the primary beam of the CBI.
We discuss these examples here.

\begin{figure}
\centerline{\psfig{file=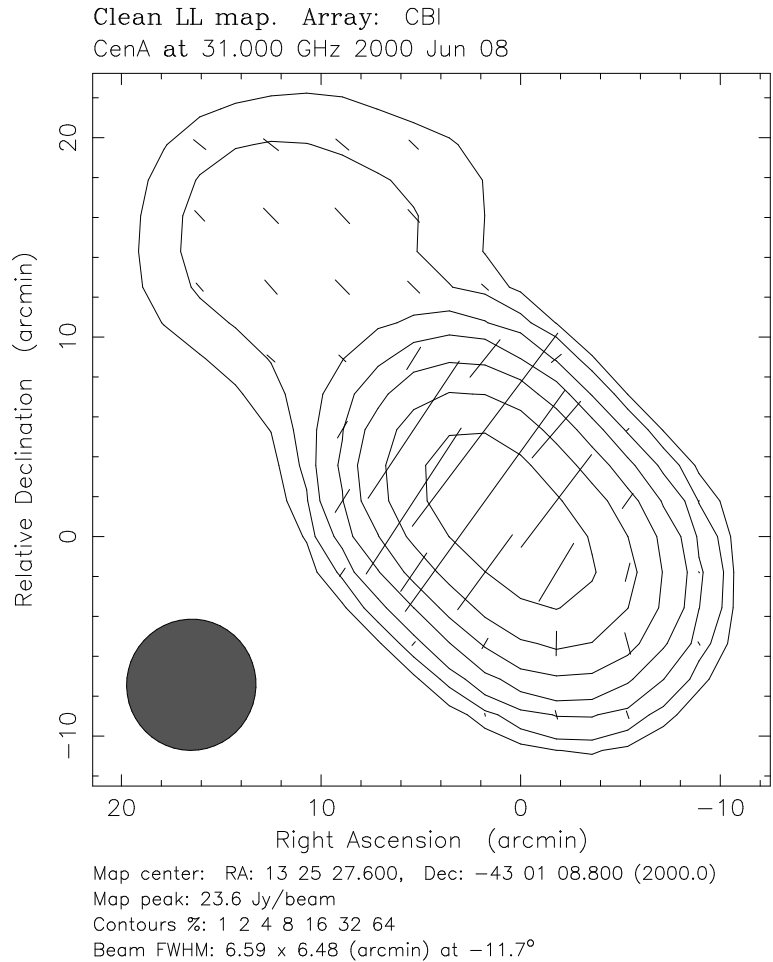,width=5.0in,angle=-90}}
\caption[Centaurus A, CBI map for {\em I} and {\em P}]{CBI map of the double inner lobes of 
Centaurus A at 31 GHz.  Contours are shown for total intensity, while the polarization 
magnitude and position angle are represented by lines.  The southern edge of the northern
middle lobe appears as the dim feature at the upper left, while the bright oval-shaped 
region of emission encompasses the northern and southern inner lobes.  The text discusses 
the distribution of polarized flux.}
\label{figure:cbi_cena}
\end{figure}

We observed Centaurus A for 6.8$^h$ with the CBI.
Figure~\ref{figure:cbi_cena} shows the
CBI map of Cen A's double inner lobes, along with the southernmost edge of the northern middle lobe.
The image is centered on the northern end of the double inner lobe, at which point the total 
intensity peaks at 20.1 Jy, the fractional polarization reaches 12\%, and the position angle is $-36^{\circ}$.
While the total intensity of the southern lobe resembles that of the northern lobe---it 
peaks at 18.7 Jy/beam---the polarization characteristics of the southern lobe are strikingly 
different;  the fractional polarization reaches 3.6\% at the total intensity peak, 
at which point the PA $\sim -37^{\circ}$.
 \cite{junkes93} present observations of the inner lobes of Cen A at 6.3 GHz with the 
Parkes 64-m telescope;
the authors report that the northern inner lobe the fractional polarization peaks at 13\%, while 
at the peak of the total intensity of the southern inner lobe the polarization rises to only $\sim 5\%$
at the southernmost edge of the lobe.
The position angle across the two inner lobes is  $-70^{\circ}< \chi< -33^{\circ}$, and it wraps around 
around to $\sim +5^{\circ}$ along the southern slope of the southern inner lobe.
The CBI results are consistent with these findings.

\begin{figure}
\centerline{\psfig{file=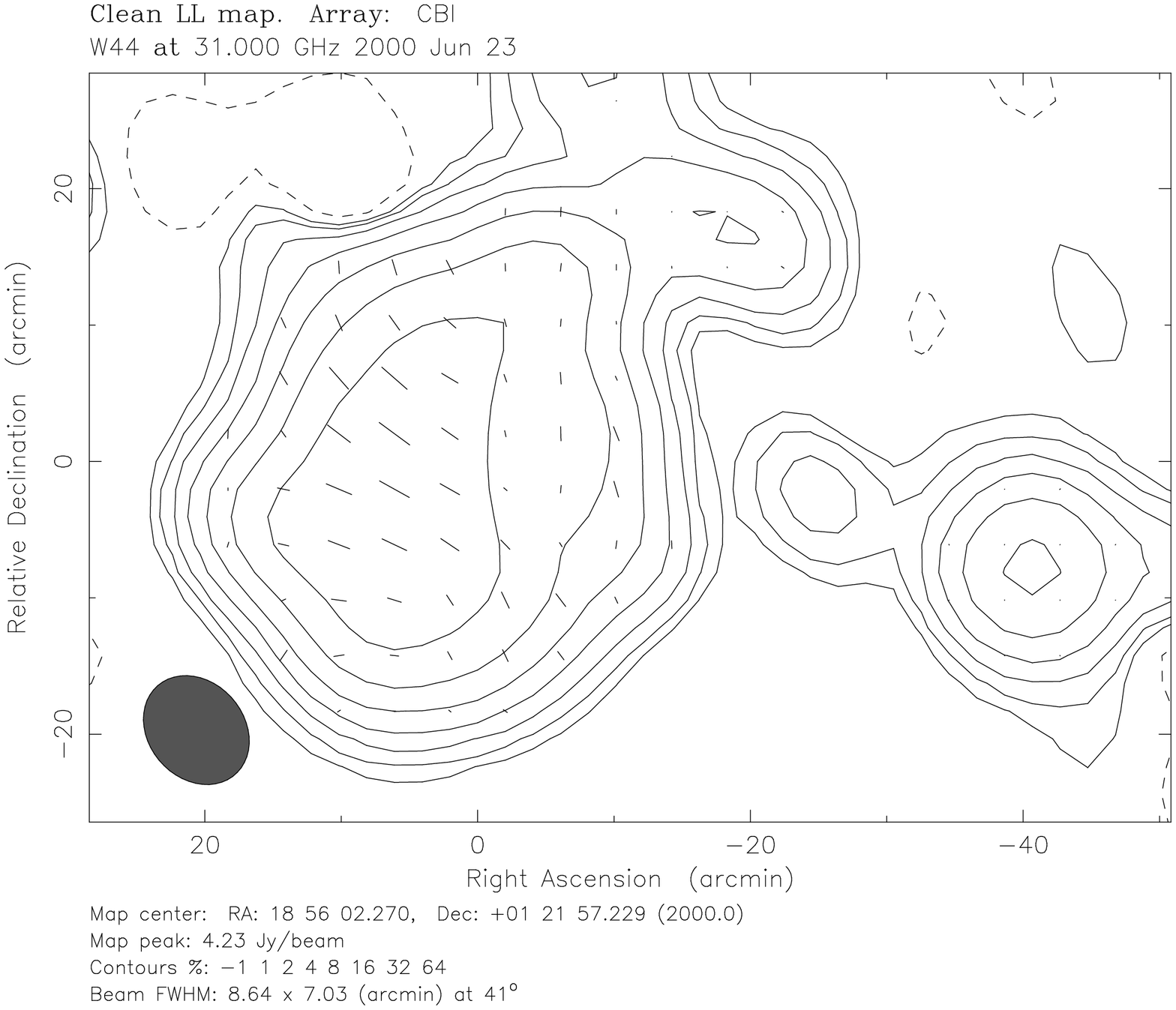,width=5.0in,angle=-90}}
\begin{singlespace}
\caption[W44, CBI map for {\em I} and {\em P}]{CBI map of supernova remnant W44 (left) and the galactic 
HII region G34.3+0.1 (right).  The HII region is unpolarized, and the lack of polarization greater 
than 0.5\% in the CBI map of G34.3+0.1 demonstrates that the CBI does not create spurious polarization greater 
than this level.}
\label{figure:w44_cbi}
\end{singlespace}
\end{figure}

We observed W44 for 2.6$^h$ with the CBI.
Figure~\ref{figure:w44_cbi} shows the CBI map of W44 after having been restored with a 
$8.6^{\prime}\times 7^{\prime}$ beam.
The remnant has a pear-shaped shell, with a distinct asymmetry arising from 
the steep density gradient in the immediate neighborhood of the remnant~\cite[]{cox99}.
The CBI maps show that the fractional polarization peaks at $\sim 33\%$ on the northwestern 
slope of the source,
and across the center of the source it is relatively uniform at 10-12\%.
While the position angle varies across the source, it is roughly uniform at $\sim 60^{\circ}$ 
across most of the emission in total intensity.
Kundu and Velusamy used the NRAO 140$^{\prime}$ telescope to map W44 at 10.7 GHz with a 3$^{\prime}$ 
beam~\cite[]{kundu72}.
The authors report that the fractional polarization peaks at $\sim 20\%$ along the NE edge, 
and it remains uniform over the dominant region of emission along the east side of the source.
At the peak of emission in total intensity the authors find that the fractional polarization 
$m\sim 20\%$.
The neighborhood of W44 contains a galactic HII region that provides a key test of
the CBI's polarization capabilities.
The emission from this source, G34.3+0.1, is due to free-free emission, so the source should be 
unpolarized.
The fractional polarization at the total intensity of the emission is $\sim 0.5\%$, so we conclude
that the CBI is not creating spurious polarization at greater than this level.
These tests gave us great confidence in the potential of the CBI for polarization observations,
and they were an important factor in our decision to upgrade the instrument to carry out a focussed 
program of polarization observations.

\section{Likelihood Analysis of the Polarization Data}

The method of maximum likelihood was used to test the data for the presence 
of a hypothetical signal.
The {\em likelihood} of the data {\bf x} given a theory {\bf q} is given by

$$\mathcal{L}({\bf x}|{\bf q}) = {1\over \pi^{N_d}|{\bf C}({\bf q})|}\textsf{exp}\Bigl[-{\bf x}^t{\bf C}^{-1}({\bf q}){\bf x}\Bigr]$$

\noindent where {\bf x} is a data vector of length $N_d$ and the covariance matrix {\bf C} quantifies
the correlations between these data for the model under test.
In this analysis, {\bf x} is a vector that contains the real and imaginary components of 
the {\em LR} visibilities $\mathcal{V}^{LR}({\bf u})$ that populate both halves of the $(u,v)$ plane.
{\bf C}({\bf q}) consists of a theoretical correlation {\bf M} and a diagonal noise matrix {\bf N}: 
${\bf C}({\bf q}) = {\bf M}({\bf q}) + {\bf N}$.
The model {\bf q} that maximizes the likelihood, or, equivalently, the {\em log}-likelihood

$$\textsf{ln}\mathcal{L}({\bf x}|{\bf q}) = -N_d\textsf{ln}[\pi] - \textsf{ln}[|{\bf C}({\bf q})|] -{\bf x}^t{\bf C}({\bf q})^{-1}{\bf x}$$

\noindent is regarded as the model that is most consistent with the data.
The model may be a function of $\ell$;  
we assume a model with {\em flat bandpower}, for which 
$q_i=\mathcal{C}^{EE}_{\ell} = \ell(\ell+1)/2\pi C^{EE}_{\ell}$ 
is constant for the $\ell$-range defined for band $i$.
Several authors \cite[e.g.,][]{hobson95,myers03} discuss techniques for 
applying the method of maximum likelihood to visibility data;  we have implemented 
aspects of the approaches discussed by these authors with the assumption that 
$\mathcal{C}^{BB}_{\ell}\equiv 0$.

The deep field observations described in this work yielded $\sim 10^7$ visibilities, each of 
which corresponds to an $8.4^s$ integration for a single baseline and channel, 
so the visibilities were averaged to reduce the covariance matrix to tractable proportions.
The visibilities were averaged in three passes.
First, the $\sim 50$ 8.4$^s$ integrations in each 8$^m$ scan were averaged to form a single visibility 
for that scan.
The uncertainty for the scan-averaged visibility was computed from the scatter in the constituent visibilities.
This procedure introduces a downward bias in the noise, so the elements of the noise covariance matrix 
{\bf N} were scaled upward by 6\% to correct for this bias \cite[]{mason02}.
Next, all of the visibilities for all nights were averaged by $(u,v)$ point, and finally, 
to truncate the size of the covariance matrix, the visibilities were averaged by band.
The band average has the potential to bias the 
best-fit bandpowers, so we analyzed two simulated sets of low signal-to-noise data, one with a single 
$\Delta\nu = 1$ GHz band centered at the middle of the CBI band, and another with the entire $\Delta\nu = 10$ GHz 
band averaged to a single $\Delta\nu = 1$ GHz band that was centered at the same frequency as the first set.
We found that the upper limits obtained from the two sets of data were consistent;  this 
should be the case, as the data are dominated by noise.
The final data set consisted of 185 and 149 discrete $(u,v)$ points for the 08$^h$ field and 
20$^h$ fields, respectively. 
To expedite the likelihood calculation, these visibilities were sorted into three bins 
based on $|{\bf u}|$;  because of 
the spacing between the antennas, this binning scheme resulted in one based physical length:  band 1
incorporated the 100 and 104 cm baselines, band 2 contained the 173 and 200 cm baselines, and band 3
contained the remaining long baselines.
The resulting upper limits do not correct for correlations between these bands.



Simulations provide insights about the effects of errors in the calibration,
so we simulated data with errors in the complex gain $G_{jk}$ and complex 
leakage $\epsilon_{jk}$ (Equation 2).
The simulations demonstrated that substantial errors in the gain phase $G_{\phi}$ 
($\delta G_{\phi}\sim 6^{\circ}$, or 10\% of a radian) result in negligible changes 
($\sim 1\%$) to the best-fit bandpower, while changes to the gain amplitude $G_A$ scale 
the best-fit bandpowers quadratically, as expected.
Systematic errors in the leakage calibration are of particular concern 
because they can mimic real polarization in the CMBR.
These simulations show that the errors in the leakage phase $\epsilon_{\phi}$ of the 
instrumental polarization do not affect the best-fit bandpowers (for fixed nonzero $\epsilon_A$), 
while errors in the leakage amplitude $\epsilon_A$ tend to {\em increase} the best-fit bandpowers regardless of 
whether they overestimate or underestimate the true leakage amplitude;  this must
be the case, since the power in fluctuations is purely additive.
Errors in $\epsilon_A$ contribute in quadrature with the intrinsic polarization on the sky:  
$\mathcal{C}^{EE}_{\ell}\rightarrow \mathcal{C}^{EE}_{\ell}+\delta\epsilon_A^2 \mathcal{C}^{TT}_{\ell}$.
A 20\% error in the amplitude of 10\% instrumental polarization, for example, 
tends to bias the amplitude of the best-fit bandpower $\sqrt{\mathcal{C}^{EE}_{\ell}}$ {\em upward} 
by less than 2\% for a generic standard cosmology.
We are therefore confident that the bandpowers reported in this work are not contaminated by errors in the 
leakage correction by more than this level.

Since we report upper limits in this paper, our primary concern is that systematic calibration 
errors do not cause us to {\em underestimate} these limits. 
The simulations demonstrated that of the four types of calibration errors ($G_A$, $G_{\phi}$, 
$\epsilon_A$, and $\epsilon_{\phi}$), only a systematic error in the gain amplitude can bias the limits downward,
and, as noted previously, a variety of cross-checks demonstrated that the error on $G_A$ is 10\%.
All of the sources of uncertainty---the assumptions for the likelihood calculation and errors in 
the instrumental polarization calibration---tend to result in overestimates of the best-fit bandpowers; 
we are confident that the limits reported herein do not underestimate the sky signal
beyond the uncertainty in the gain calibration.

\section{Results}

The 278$^h$ of deep field data yielded several upper limits on $\mathcal{C}^{EE}_{\ell}$.
Table~\ref{table:cbi_likelihood_results} lists the 95\% c.l. results for the measurements 
of the two fields and the joint fit to the fields;  these were obtained by integrating the likelihood 
from $q\equiv 0$.
We have assumed that $\mathcal{C}^{BB}_{\ell}\equiv 0$.
For each band, the band center is the peak of the summed diagonal elements of ${\bf M}$, while 
the band width is the FWHM of the summed diagonal elements of ${\bf M}$. 
As a cross-check, the likelihood routine was modified to address $\mathcal{C}^{TT}_{\ell}$;
it was tested on the short-baseline 08$^h$ field data, for which it yielded 
$\sqrt{\mathcal{C}^{TT}_{\ell}}=66.8^{+14.1}_{-11.1}$ $\mu$K.
This value is consistent with the value obtained by the CBI for nearly the same $\ell$-range:
$\sqrt{\mathcal{C}^{TT}_{\ell}}=62.9^{+11.3}_{-7.9}$ $\mu$K \cite[]{padin02};  the two sets of data 
have differing $(u,v)$ coverage, so the two measurements are not identical.

These upper limits are consistent with the rapidly burgeoning body of CMBR polarization data.
The limit at $\ell=603$ is consistent with limits in the same region from DASI and CBI.
The limits for the higher $\ell$ bins are consistent with predictions for $\mathcal{C}^{EE}$ for 
generally accepted families of models.
The limits also provide constraints on polarized emission from galactic synchrotron emission 
and polarized point sources in these regions and on these scales.
This poineering polarization effort with the CBI provided great confidence in the polarization 
capabilities of the instrument, and it was a central consideration in our decision to upgrade the 
CBI for a dedicated polarization program.



\section{Acknowledgements}
We gratefully acknowledge CBI Project Scientist Steve Padin for his contributions as 
chief architect of the instrument.
We thank Brian Mason and Patricia Udomprasert for their help with the observations.
We are grateful to Steve Myers and Carlo Contaldi for suggestions about the likelihood calculation.
We thank Barbara and Stanley Rawn, Jr., for their continuing support of the CBI project.
This work was supported by the National Science Foundation under grants AST 9413935, 9802989, 0098734,
and 0206412.
JKC acknowledges support from NSF grant OPP-0094541 to the Kavli Institute for Cosmological Physics.

Facilities:  \facility{VLA}.

\clearpage

\begin{table}
\begin{center}
\caption{Gaussian model components for Tau A at 31 GHz}
\medskip
\label{table:taua_models}
\begin{tabular}{ccccccc}
\tableline\tableline
  & $S_{\nu}$ & $x_0$ & {$y_0$\tablenotemark{a}} & $\sigma$ & $b/a$ & {$\phi$\tablenotemark{b}} \\
  Component &  (Jy) &  ($^{\prime\prime}$) & ($^{\prime\prime}$) &  ($^{\prime}$) &  & ($^{\circ}$) \\\hline
  I  	&  355.3 & 0.0   &  0.0		& 3.58	&  0.66	&  -50	 \\
  Q  	&  14.9 & -48.8  &  116.9	& 2.93	&  0	&   83	 \\
  U  	& -23.9 & -30.1  &  128.2 	& 2.28	&  0.52	&   56	 \\
\hline
\end{tabular}

\tablenotetext{a}{$x_0$ and $y_0$ are positions of the centroids of the model components, measured 
with respect to that for the total intensity.}
\tablenotetext{b}{$\sigma$, $b/a$, and $\phi$ are the major axis width, axial ratio, and orientation 
of the elliptical Gaussian model to which each component was fit.}
\end{center}
\end{table}

\clearpage

\begin{table}
\begin{center}
\caption{Upper limits on $\mathcal{C}^{EE}_{\ell}$, 95\% confidence}  
\medskip
\label{table:cbi_likelihood_results}
\begin{tabular}{ccccccc}
\hline\hline
 & &  &  & {08$^h$} ${\sqrt{q}}$ & {20$^h$} $\sqrt{q}$  & joint $\sqrt{q}$ \\
Band & $\ell_{min}$ & $\ell_c$ & $\ell_{max}$ & ($\mu$K) & ($\mu$K) & ($\mu$K) \\\hline
1 &\  446  &\  603 &\  779  & 14.1 &\ 8.1  &\  7.0 \\
2 &\  930  &  1144 &  1395  & 21.2 & 15.9  &  12.8 \\
3 &  1539  &  2048 &  2702  & 45.3 & 27.7  &  25.1  \\\hline

\end{tabular}
\end{center}
\end{table}

\clearpage

\clearpage

\clearpage

\begin{thebibliography}
\expandafter\ifx\csname natexlab\endcsname\relax\def\natexlab#1{#1}\fi

\bibitem[{{Barkats} {et~al.}(2004)}]{barkats04}
        {Barkats}, D., {Bischoff}, C., {Farese}, P., {Fitzpatrick}, L., {Gaier}, T.,
	{Gunderson}, J., {Hedman}, M., {Hyatt}, L., {McMahon}, J., {Samtleben}, D.,
        {Staggs}, S., {Vanderlinde}, K., \& {Winstein}, B.
	2004, ApJ, submitted (astro-ph/0409380)

\bibitem[{{Bennett} {et~al.}(2003)}]{bennett03}
 	{Bennett}, C., {Halpern}, M., {Hinshaw}, G. , {Jarosik}, N., {Kogut}, A., {Limon}m M., {Meyer}, S., 
	{Page}, L., {Spergel}, D., {Tucker}, G., {Wollack}, E., {Wright}, E., {Barnes}, C., {Greason}, M., 
	{Hill}, R., {Komatsu}, E., {Nolta}, M., {Odegard}, N., {Peirs}, H., {Verde}, L., \& {Weiland}, J.
	2003, ApJS, 148, 1

\bibitem[{{Cartwright}(2003)}]{cartwright03}
	{Cartwright}, J.~K.
	2003, Ph.D Thesis, California Institute of Technology

\bibitem[{{Conway} \& {Kronberg}(1969)}]{conway69}
	{Conway}, R.~G. \& {Kronberg}, P.~P.
	1969, MNRAS, 142, 11

\bibitem[{{Cox} {et~al.}(1999)}]{cox99}
    	{Cox}, D.~P., {Shelton}, R.~L., {Maciejewski}, W., {Smith}, R.~K., {Plewa}, T., 
	{Pawl}, A., \& {R{\' o}{\. z}yczka}, M.
	1999, ApJ, 524, 179

\bibitem[{{Delabrouille} {et~al.}(2002)}]{delabrouille02}
 	{Delabrouille}, J., {Kaplan}, J., \& {The Planck HFI Consortium}.
	2002, in AIP Conf. Proc. 609, Astrophysical Polarized Backgrounds, ed. 
	S. Cecchini, S. Cortiglioni, R. Sault, \& C. Sbarra (New York:  AIP), 135

\bibitem[{{Halverson} {et~al.}(2002)}]{halverson02}
   	{Halverson}, N., {Leitch}, E., {Pryke}, C., {Kovac}, J.,
        {Carlstrom}, J., {Holzapfel}, W.,  {Dragovan}, M.,
        {Cartwright}, J.,  {Mason}, B.,  {Padin}, S.  {Pearson}, T.,  
        {Readhead}, A., \& {Shepherd}, M.
	2002, ApJ, 568, 38

\bibitem[{{Haslam} {et~al.}(1982)}]{haslam82}
    	{Haslam}, C.~G.~T., {Stoffel}, H., {Salter}, C.~J., \& 
        {Wilson}, W.~E.
	1982, A\&AS, 47, 1

\bibitem[{{Hedman} {et~al.}(2001)}]{hedman01}
 	{Hedman}, M.~M., {Barkats}, D., {Gundersen}, J.~O., 
        {Staggs}, S.~T., \& {Winstein}, B.
	2001, 548, L111

\bibitem[{{Hedman} {et~al.}(2002)}]{hedman02}
 	{Hedman}, M.~M., {Barkats}, D., {Gundersen}, J.~O., 
        {McMahon}, J.~J., {Staggs}, S.~T., \& {Winstein}, B.
	2002, ApJL, 573, L73

\bibitem[{{Hinshaw} {et~al.}(2003)}]{hinshaw03}
	{Hinshaw}, G., {Spergel}, D., {Verde}, L., {Hill}, R., {Meyer}, S., {Barnes}, C., 
	{Bennett}, C., {Halpern}, M., {Jarosik}, N., {Kogut}, A., {Komatsu}, E., {Limon}, M., 
	{Page}, L., {Tucker}, G., {Weil}, J., {Wollack}, E., \& {Wright}, E.
	2003, ApJS, 148, 135

\bibitem[{{Hobson} {et~al.}(1995)}]{hobson95}
 	{Hobson}, M.~P., {Lasenby}, A.~N., \& {Jones}, M.
	1995, MNRAS, 275, 863

\bibitem[{{Hu} \& {Dodelson}(2002)}]{hu02}
    	{Hu}, W. \& {Dodelson}, S.
 	2002, ARAA, 40, 171

\bibitem[{{Johnson} {et~al.}(2003)}]{johnson03}
	{Johnson}, B.~R., {Abroe}, M.~E., {Ade}, P., {Bock}, J., {Borrill}, J., {Collins}, J.~S., 
	{Ferreira}, P., {Hanany}, S., {Jaffe}, A.~H., {Jones}, T., {Lee}, A.~T., {Levinson}, L., 
	{Matsumura}, T., {Rabii}, B., {Renbarger}, T., {Richards}, P.~L., {Smoot}, G.~F., {Stompor}, R., 
	{Tran}, H.~T., \& {Winant}, C.~D.
	2003, in The Cosmic Microwave Background and its Polarization, ed. S. Hanany \& K.A. Olive 
	(Amsterdam:  Elsevier), in press (astro-ph/0308259)

\bibitem[{{Junkes} {et~al.}(1993)}]{junkes93}
        {Junkes}, N. , {Haynes}, R.~F., {Harnett}, J.~I., \& {Jauncey}, D.~L.
        1993, {A\&A}, 269, 29

\bibitem[{{Kamionkowski} \& {Kosowsky}(1999)}]{kamionkowski99}
      	{Kamionkowski}, W. \& {Kosowsky}, M.
	1999, Ann. Rev. Nucl. Part. Sci., 2, 77,

\bibitem[{{Kamionkowski} {et~al.}(1997)}]{kamionkowski97}
 	{Kamionkowski}, M. {Kosowsky}, A., \& {Stebbins}, A.
	1997, Phys. Rev. D, 55, 7368

\bibitem[{{Keating} {et~al.}(2001)}]{keating01}
 	{Keating}, B.~G., {O'Dell}, C.~W., {de Oliveira-Costa}, A., 
        {Klawikowski}, S., {Stebor}, N., {Piccirillo}, L., 
        {Tegmark}, M., \& {Timbie}, P.~T.
	2001, ApJL, 560, L1

\bibitem[{{Kogut} {et~al.}(2003)}]{kogut03}
        {Kogut}, A., {Spergel}, D.~N., {Barnes}, C., {Bennett}, C.~L., {Halpern}, M., {Hinshaw}, G., 
	{Jarosik}, N., {Limon}, M.~S., {Meyer}, S., {Page}. L., {Tucker}, G., {Wollack}, E., \& {Wright}, E.~L.
	2003, ApJS, 148, 161

\bibitem[{{Kovac} {et~al.}(2002)}]{kovac02}
     	{Kovac}, J., {Leitch}, E., {Pryke}, C., {Carlstrom}, J., 
     	{Halverson}, N., \& {Holzapfel}, W.~L.
	2002, Nature, 420, 772

\bibitem[{{Kuo et~al.}(2002)}]{kuo02}
	{Kuo}, C., {Ade}, P., {Bock}, J., {Cantalupo}, C., {Daub}, M., {Goldstein}, J., 
	{Holzapfel}, W., {Lange}, A., {Lueker}, M., {Newcomb}, M., {Peterson}, J., 
	{Ruhl}, J., {Runyan}, M., {Torbet}, E., 
	(2004), ApJ, 600, 32

\bibitem[{{Leitch} {et~al.}(2004)}]{leitch04}
        {Leitch}, E., {Kovac}, J., {Halverson}, N., {Carlstrom}, J., {Pryke}, C., 
	\& {Smith}, M.
	2004, ApJ, submitted (astro-ph/0409357)

\bibitem[{{Leitch} {et~al.}(2002)}]{leitch02}
        {Leitch}, E., {Kovac}, J., {Pryke}, C., {Reddall}, B., {Sandberg}, E., {Dragovan}, M., 
	{Carlstrom}, J., {Halverson}, N., \& {Holzapfel}, W.
	2004, Nature, 420, 763

\bibitem[{{Kundu} \& {Velusamy}(1972)}]{kundu72} 
	{Kundu}, M.~R. \& {Velusamy}, T.
	1972, A\&A, 20, 237

\bibitem[{{Masi} {et~al.}(2002)}]{masi02}
	{Masi}, S., {Ade}, P.~A.~R., {Bock}, J.~J., {Boscaleri}, A., 
        {de Bernardis}, P., {de Troia}, G., {di Stefano}, G., 
        {Hristov}, V.~V., {Iacoangeli}, A., {Jones}, W.~C., 
        {Kisner}, T., {Lange}, A.~E., {Mauskopf}, P.~D., {Mac Tavish}, C., 
        {Montroy}, T., {Netterfield}, C.~B., {Pascale}, E., 
        {Piacentini}, F., {Pongetti}, F., {Romeo}, G., {Ruhl}, J.~E., 
        {Torbet}, E., \& {Watt}, J.
	2002, in AIP Conf. Proc. 609, Astrophysical Polarized Backgrounds, ed. 
	S. Cecchini, S. Cortiglioni, R. Sault, \& C. Sbarra (New York:  AIP), 122

\bibitem[{{Mason} {et~al.}(2003)}]{mason02}
     	{Mason}, B.~S., {Pearson}, T.~J., {Readhead}, A.~C.~S., {Shepherd}, M.~C., 
	{Sievers}, J.~L., {Udomprasert}, P.~S., {Cartwright}, J,~K., {Farmer}, A.~J., {Padin}, S., 
	{Myers}, S.~T., {Bond}, J.~R., {Contaldi}, C.~R., {Pen}, U.-L., {Prunet}, S., {Pogosyan}, D., 
	{Carlstrom}, J.~E., {Kovac}, J., {Leitch}, E.~M., {Pryke}, C., {Halverson}, N.~W., 
	{Holzapfel}, W.~L., {Altamirano}, P., {Bronfman}, L., {Casassus}, S., {May}. J., \& {Joy}, M.
	2003, ApJ, 591, 540

\bibitem[{{Myers} {et~al.}(2003)}]{myers03}
 	{Myers}, S.~T., {Contaldi}, C.~R., {Bond}, J.~R., {Pen}, U.-L., {Pogosyan}, D., 
	{Prunet}, S., {Sievers}, J.~L., {Mason}, B.~S., {Pearson}, T.~J., {Readhead}, A.~C.~S., \& 
	{Shepherd}, M.~C.
	2003, ApJ, 591, 575

\bibitem[{{Netterfield} {et~al.}(2002)}]{netterfield02}
 	{Netterfield}, C.,  {Ade}, P.,  {Bock}, J.,  
        {Bond}, J., {Borrill}, J.,  {Boscaleri}, A.,  {Coble}, K. , 
        {Contaldi}, C., {Crill}, B., {de Bernardis}, P.,  
        {Farese}, P., {Ganga}, K., {Giacometti}, M. {Hivon}, E., 
        {Hristov}, V., {Iacoangeli}, A., {Jaffe}, A.,  
        {Jones}, W., {Lange}, A., {Martinis}, L. {Masi}, S., 
        {Mason}, P., {Mauskopf}, P., {Melchiorri}, A. {Montroy}, T.,  
        {Pascale}, E., {Piacentini}, F. {Pogosyan}, D. {Pongetti}, F.,  
        {Prunet}, S.,{Romeo}, G., {Ruhl}, J., \& {Scaramuzzi}, F.
	2002, ApJ, 571, 604

\bibitem[{{Netterfield} {et~al.}(1997)}]{netterfield97}
	{Netterfield}, C.~B., {Devlin}, M.~J., {Jarolik}, N., 
        {Page}, L., \& {Wollack}, E.~J.
	1997, ApJ, 474, 47	

\bibitem[{{Padin} {et~al.}(2001)}]{padin01}
 	{Padin}, S., {Cartwright}, J.~K., {Mason}, B.~S., {Pearson}, T.~J., 
        {Readhead}, A.~C.~S., {Shepherd}, M.~C., {Sievers}, J., 
        {Udomprasert}, P.~S., {Holzapfel}, W.~L., {Myers}, S.~T., 
        {Carlstrom}, J.~E., {Leitch}, E.~M., {Joy}, M., {Bronfman}, L., \&
        {May}, J.
	2001, ApJL, 549, L1

\bibitem[{{Padin} {et~al.}(2002)}]{padin02}
	{Padin}, S., {Shepherd}, M.~C., {Cartwright}, J.~K., 
        {Keeney}, R.~G., {Mason}, B.~S., {Pearson}, T.~J., 
        {Readhead}, A.~C.~S., {Schaal}, W.~A., {Sievers}, J., 
        {Udomprasert}, P.~S., {Yamasaki}, J.~K., {Holzapfel}, W.~L., 
        {Carlstrom}, J.~E., {Joy}, M., {Myers}, S.~T., \& {Otarola}, A.
	2002, PASP, 114, 83

\bibitem[{{Pearson} {et~al.}(2003)}]{pearson02}
 	{Pearson}, T.~J., {Mason}, B.~S., {Readhead}, A.~C.~S.,
	{Shepherd}, M.~C., {Sievers}, J.~L., {Udomprasert}, P.~S., 
	{Cartwright}, J.~K., {Farmer}, A.~J., {Padin}, S., {Myers}, S.~T., 
	{Bond}, J.~R., {Contaldi}, C.~R., {Pen}, U.-L., {Prunet}, S., 
	{Pogosyan}, D., {Carlstrom}, J.~E., {Kovac}, J., {Leitch}, E.~M., 
	{Pryke}, C., {Halverson}, N.~W., {Holzapfel}, W.~L., {Altamirano}, P., 
	{Bronfman}, L., {Casassus}, S., {May}, J., \& {Joy}, M.
	2003, ApJ, 591, 556

\bibitem[{{Piccirillo} {et~al.}(2002)}]{piccirillo02}
	{Piccirillo}, L., {Ade}, P.~A.~R., {Bock}, J.~J., {Bowden}, M., 
        {Church}, S.~W., {Ganga}, K., {Gear}, W.~K., {Hinderks}, J., 
        {Keating}, B.~G., {Lange}, A.~E., {Maffei}, B., {Malli{\' e}}, O., 
        {Melhuish}, S.~J., {Murphy}, J.~A., {Pisano}, G., {Rusholme}, B., 
        {Taylor}, A., \& {Thompson}, K.
	2002, in AIP Conf. Proc. 609, Astrophysical Polarized Backgrounds, ed. 
	S. Cecchini, S. Cortiglioni, R. Sault, \& C. Sbarra (New York:  AIP), 159

\bibitem[{{Farese} {et~al.}(2003)}]{farese03}
	{Farese}, P.,  {Dall'Oglio}, G.,  {Gundersen}, J., {Keating}, B., 
	{Klawikowski}, S., {Knox}, L., {Levy}, A., {Lubin}, P., {O'Dell}, C., 
	{Peel}, A., {Piccirillo}, L., {Ruhl}, J. \& {Timbie}, P.
	2003, ApJ, submitted (astro-ph/0308309)


\bibitem[{{Readhead} {et~al.}(2004a)}]{readhead04a}
 	{Readhead}, A.~C.~S., {Mason}, B.~S., {Contaldi}, C.~R., 
	{Pearson}, T.~J., {Bond}, J.~R., {Myers}, S.~T., {Padin}, S., 
	{Sievers}, J.~L., {Cartwright}, J.~K., {Shepherd}, M., 
	{Pogosyan}, D., {Prunet}, S., {Altamirano}, P., 
	{Bustos}, R., {Bronfman}, L., {Casassus}, S., {Holzapfel}, W.~L., 
	{May}, J., {Pen}, U.-L., {Torres}, S., \& {Udomprasert}, P.~S.
	2004, ApJ, 609, 498


\bibitem[{{Readhead} {et~al.}(2004b)}]{readhead04b}
 	{Readhead}, A.~C.~S., {Myers}, S.~T., {Pearson}, T.~J.,
	{Sievers}, J.~L., {Mason}, B.~S., {Contaldi}, C.~R., 
	{Bond}, J.~R., {Altamirano}, P., {Bustos}, R., 
	{Achermann}, C., {Bronfman}, L., {Carlstrom}, J., 
	{Cartwright}, J.~K., {Casassus}, S., {Dickinson}, C.,
	{Kovac}, J., {Leitch}, E., {May}, J., {Padin}, S., 
	{Pogosyan}, D., {Pospieszalski}, M., {Pryke}, C., 
	{Reeves}, R., {Shepherd}, M., \& {Torres}, S.
	2004, Science, in press, (astro-ph/0409569)

\bibitem[{{Subrahmanyan} {et~al.}(2002)}]{subrahmanyan00}
 	{Subrahmanyan}, R., {Kesteven}, M.~J., {Ekers}, R.~D., {Sinclair}, M., \& {Silk}, J.
        2002, MNRAS, in press (astro-ph/0002467)

\bibitem[{{Villa} {et~al.}(2002)}]{villa02}
    	{Villa}, F., {Mandolesi}, N., {Bersanelli}, M., {Butler}, R.~C., 
        {Burigana}, C., {Mennella}, A., {Morgante}, G., {Sandri}, M., \& the 
        LFI Consortium.
	2002, in AIP Conf. Proc. 609, Astrophysical Polarized Backgrounds, ed. 
	S. Cecchini, S. Cortiglioni, R. Sault, \& C. Sbarra (New York:  AIP), 144

\bibitem[{{Wollack} {et~al.}(1993)}]{wollack93}
 	{Wollack}, E.~J., {Jarosik}, N.~C., {Netterfield}, C.~B., 
        {Page}, L.~A., \& {Wilkinson}, D.
	1993, ApJL, 419, L49	

\end{thebibliography}
\end{document}